\documentclass[10pt,conference]{IEEEtran}
\usepackage{fancyhdr}
\fancypagestyle{IEEEtitlepagestyle}{
  \fancyhf{}
  \fancyfoot[C]{\sffamily\footnotesize\thepage}
  \fancyfoot[L]{\sffamily\scriptsize Copyright held by the owner/author(s). Presented at\\
3rd International Workshop on Overlay Architectures for FPGAs (OLAF2017)
\\Monterey, CA, USA, Feb. 22, 2017.}

}
\def\TITLE{\huge Out-of-Order Dataflow Scheduling for FPGA Overlays}

\usepackage{amsmath}                
\usepackage{amssymb}                
\usepackage{amsfonts}               
\usepackage{balance}                
\usepackage{mathtools}
\usepackage[justification=centering]{caption} 
\usepackage{graphicx}               
\usepackage[bookmarks=false]{hyperref} 
\usepackage{multirow}               
\usepackage{booktabs}               

\def\ie{\emph{i.e.~}}
\def\eg{\emph{e.g.~}}

\DeclarePairedDelimiter{\ceil}{\lceil}{\rceil}
\def\tilde{{\raise.17ex\hbox{$\scriptstyle\mathtt{\sim}$}}}

\begin{document}



\author{
 \IEEEauthorblockN{Siddhartha}
 \IEEEauthorblockA{Nanyang Technological University, Singapore\\
 siddhart005@e.ntu.edu.sg}
\and
 \IEEEauthorblockN{Nachiket Kapre}
 \IEEEauthorblockA{University of Waterloo, Canada\\
 nachiket@uwaterloo.ca}
}
\title{\TITLE}

\date{}

\maketitle
\begin{abstract}
We exploit floating-point DSPs in the Arria10 FPGA and multi-pumping feature of
the M20K RAMs to build a dataflow-driven soft processor fabric for large graph
workloads. In this paper, we introduce the idea of out-of-order node
scheduling across a large number of local nodes (thousands) per processor by
combining an efficient node tagging scheme along with leading-one detector
circuits. We use a static one-time node labeling algorithm to sort nodes based
on criticality to organize local memory inside each soft processor.  This
translates to a small $\approx$6\% memory overhead. When compared to a memory-expensive FIFO-based first-come-first-serve approach used in previous studies, we deliver up to 50\% performance improvement while eliminating the cost of the FIFOs. On the Arria10 10AX115S board, we can create an overlay
design of up to 300 processors connected by high bandwidth Hoplite NoC at
frequencies up to 250\,MHz. 
\end{abstract}

\section{Introduction}

We see a rising adoption of modern FPGA technology into existing cloud computing frameworks -- recent notable developments include the Microsoft Bing Catapult project and the Amazon EC2 F1 instance. The Catapult project demonstrated that FPGAs as coprocessors/accelerators can deliver better throughput and energy efficiency on Bing search engine workloads when compared to general purpose computers. The Amazon EC2 F1 instance heralds a new future for FPGA-based cloud computing through commoditization of access. Nevertheless, FPGA-based development remains challenging due to the long development and debug cycles. Overlay architectures address this challenge by abstracting away complex and tedious RTL workflow into specialized soft-processors designed to accelerate specific workloads. The overlay architectures lower the programmability barrier, while leveraging on the computing power and low cost of FPGAs. In this extended abstract, we showcase the preliminary results for an improved token dataflow soft processor overlay designed to tackle floating-point graph workloads. Our overlay targets the Arria 10 FPGA, which allows us to take advantage of the hardened floating-point DSP block to save logic resources.

A token dataflow processor (TDP) executes instructions based on a simple dataflow-firing rule: instructions are executed only when the actor/node in the dataflow graph has received all its inputs/operands. There are no program counters and execution of instructions can happen in parallel asynchronously at each processing element (PE). Hence, token dataflow architectures implicitly expose any parallelism available in the graph, and are especially useful when working on sparse data workloads that are riddled with irregular memory access patterns, \eg indirect pointer addressing. Most TDP designs are based on the MIT Static Dataflow~\cite{dennis_dataflow_sigarch1975} machine. 
In this work, we propose two key improvements to the original TDP design: (1) more efficient on-chip block RAMs (BRAMs) utilization, and (2) a cheap out-of-order scheduling circuit for improved performance on very large workloads.

The state-of-the-art TDPs are in-order processors, \ie packets are processed (and subsequently generated) in the order they arrive at a PE. While this is cheap to implement, it can have detrimental effects on performance as nodes in a dataflow graph can have varying importance to the completion of the graph evaluation that does not follow arrival order. 
The other, more subtle, downside of in-order processors is the resource wastage for buffering the queue of ready nodes inside a PE, typically implemented as an FPGA BRAM-based FIFO. To ensure deadlock-free operation, the FIFO depth has to assume the worst-case scenario, which quickly becomes intractable for very large overlay sizes. In contrast, an out-of-order approach should deliver better performance, but is plagued by potentially large logic overheads for large graph sizes. Modern out-of-order CPU cores spend a large fraction of their die area to out-of-order logic that tracks dependencies. Luckily, dataflow overlay node operations have their dependencies encoded explicitly in the graph structure and do not need to be discovered on-the-fly as needed by CPU instructions. 
Our improved out-of-order TDP tags each node with extra criticality flags, and uses a hierarchical scheme to determine node execution order. This introduces a small $\approx$6\% overhead in memory but frees up the expensive buffer FIFO BRAMs, which can now be used to store larger graphs.

Eventually, we deliver a scalable token dataflow overlay architecture in which PEs communicate with each other using a lightweight, high-bandwidth 56b-wide Hoplite~\cite{kapre_hoplite_fpl2015} router. The PEs and routers are arranged in a 2D torus topology. In the following sections, we highlight the principles behind the out-of-order scheduler, and showcase preliminary simulation results on graph workloads.

\section{Token Dataflow Processor Design}

\subsection{Datapath}
The TDP can accept one packet from the network every cycle. Once a valid packet is received at the inputs, the processor unpacks the packet and determines if the target node has received all its inputs. If the node is ready to ``fire'', the operands are sent to the ALU (with the appropriate instruction opcode). Otherwise, the packet's payload is stored in the local graph memory. Similarly, computed output from the ALU is stored in the graph memory, and the node is flagged as a ``ready node'' that can now be processed by the packet generation logic. The packet generation logic is a non-deterministic multi-cycle process: (1) nodes can have multiple fanouts, and (2) the network may be congested. Hence, there are likely to be multiple ready nodes waiting to be processed by the packet generation logic. This is the key scheduling challenge addressed in this work.

\subsection{In-order vs Out-of-order Scheduler}
Instead of using a FIFO to buffer a queue of ready nodes, we can use a \texttt{RDY} bit-flag for each node to indicate when a node is ready for fanout processing. When the local state of a node has been computed and stored in memory by the ALU, the \texttt{RDY} bit flag is set to 1. However, because of the size of the M20K BRAMs on the Arria 10 FPGA, it is not feasible to store the bit-flag vectors of {\bf all} local nodes in {\it distributed memory}. The BRAMs (20Kb) are setup in the 512x40b configuration, hence, we can store up to 40x1b flags in each BRAM memory address. For simpler arithmetic, we use only 32b out of the 40b to store bit-flags. Also, to avoid data corruption, we need \texttt{RDY} bit-flags to indicate if all fanouts of a node have been sent as packets into the network. Therefore, for a single BRAM with 512 address locations, we need $2*\ceil*{\frac{512}{32}} = 32$ memory locations for storing all \texttt{RDY} bit-vectors ($\approx$6\% overhead). Since our TDP design is composed of 8 BRAMs/processor, we use 256x40b memory locations to store all bit-flags.

As there are 256x32b bit-flag memory locations, we must scan, in the worst-case scenario, 256 memory locations to identify a ready node. This is a significant performance overhead, which we mitigate by designing a hierarchical scheduling strategy that utilizes a leading-ones detector (LOD) circuit. The LOD is a well-known cheap combinatorial circuit that can identify the position of the leading one in an input bit-vector. In a hierarchical scheduler design, we have an {\it OuterLOD} and an {\it InnerLOD}. The OuterLOD takes an input 128b vector (stored in distributed memory) and based on the position of the leading one, a 32b-vector stored in graph memory is retrieved. The 32b-InnerLOD can then identify the leading one in that 32b-vector, which identifies the next ready node for fanout processing. In practice, this is a deterministic 2-cycle process as opposed to the na{\"i}ve non-deterministic memory scan implementation.

The above scheduling strategy can be further improved with a static heuristic-based memory organization. Before execution, we do a one-time software criticality evaluation on the application dataflow graph. At the end of this routine, each node is labeled with a criticality parameter that indicates its importance to the computation. We use this metric to organize the graph memory contents such that nodes are placed in each local graph memory in {\bf decreasing} criticality order. When combined with hierarchical LOD-based scheduling, the processor implicitly picks the {\bf most critical node} during each scheduling pass.

\subsection{Other Optimizations}

We multipump our BRAMs to create additional virtual read/write ports. The graph structure is carefully encoded in order to maximize every bit of on-chip BRAM capacity. We synthesize two hard floating-point DSP blocks inside each PE, configured to ADD and MULTIPLY mode respectively. To meet our operating frequency targets, the DSP blocks are configured in a single-stage pipeline mode. Finally, the packet generation logic is capable of injecting one packet every cycle into the network (subject to congestion).

\section{Overlay Performance}

\begin{table}[b]
\vspace{-0.1in}
\centering
\caption{Resource utilization}
\label{resources.tbl}

\scriptsize
\begin{tabular}{c c c c c c}
\toprule
{\bf Size} & {\bf ALMs} & {\bf REGs} & {\bf DSPs} & {\bf BRAMs} & {\bf Freq.} \\
\midrule
    1		&	1.4K (0.3\%)	&	2.2K (0.1\%)	&	2  (0.1\%)	&	8 (0.3\%)	&	306 MHz\\
    256		&	367K (86\%)	&	559K (25\%)	&	512  (34\%)	&	2K (75\%)	&	258 MHz\\
\bottomrule
\multicolumn{6}{l}{\scriptsize *one Hoplite router consumes 130 ALMs, 350 registers, and runs at $>$400MHz}
\end{tabular}
\end{table}

\begin{figure}[t] \centering
\includegraphics[width=2.5in]{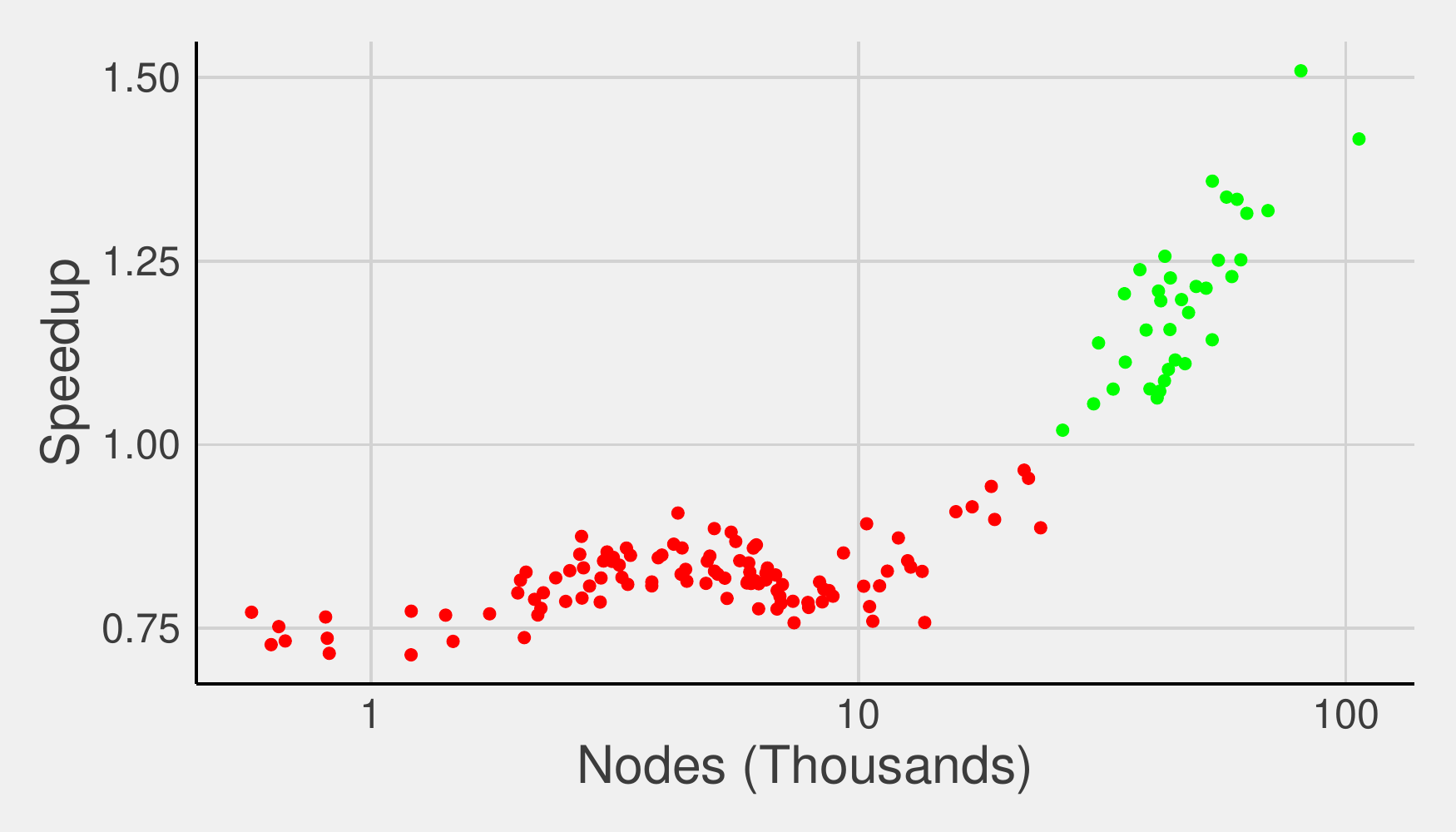}
\caption{Out-of-order scheduling: speedup over in-order scheduling vs size of dataflow graph}
\label{ooo_vs_io.fig}
\vspace{-0.2in}
\end{figure}
Table~\ref{resources.tbl} shows the resource utilization of the overlay design. The preliminary simulation experiments are conducted on dataflow graphs extracted from sparse matrix factorization kernels. The graph sizes range from a few hundred nodes/edges to $>$100K nodes/edges. We ran simulations on these workloads with overlay sizes ranging from a single PE (1x1) to 256 PEs (16x16). Figure ~\ref{ooo_vs_io.fig} shows speedups of out-of-order over in-order scheduling. At 30K and beyond application graph sizes, the in-order overlay has exhausted all available parallelism, \ie using the maximum available 256 PEs. Out-of-order schedulers are able to deliver improved performance of up to 50\% beyond this point. For larger graphs, computing along the critical path is also more important, as demonstrated by the improving speedups as graph sizes continue increasing.

Finally, the freeing up of BRAMs used for FIFOs also improves scalability of design. The 256 FIFO-based processors overlay is only capable of storing application graphs of size $\approx$100K nodes and edges, whereas the 256 PE out-of-order design can support $\approx$5$\times$ larger input graphs.

\addcontentsline{toc}{section}{References}

\bibliographystyle{abbrv}
\bibliography{library}

\end{document}